\documentclass[twocolumn]{aastex631}

\accepted{15th November 2021}

\shorttitle{Non-periodic type I Be/X-ray binary outbursts}
\shortauthors{R. G. Martin \& A. Franchini}
\begin{document}

\title{Non-periodic type I Be/X-ray binary outbursts}

\author{Rebecca G. Martin}
\affil{Department of Physics and Astronomy, University of Nevada, Las Vegas,
4505 South Maryland Parkway, Las Vegas, NV 89154, USA}
\author{Alessia Franchini}
\affiliation{Dipartimento di Fisica ``G. Occhialini", Universit\'a degli Studi di Milano-Bicocca, Piazza della Scienza 3, 20126 Milano, Italy}

%\author[0000-0002-0786-7307]{Greg J. Schwarz}
%\affil{American Astronomical Society \\
%2000 Florida Ave., NW, Suite 300 \\
%Washington, DC 20009-1231, USA}

%\author{August Muench}
%\affiliation{American Astronomical Society \\
%2000 Florida Ave., NW, Suite 300 \\
%Washington, DC 20009-1231, USA}
%\collaboration{(AAS Journals Data Scientists collaboration)}

%% Note that the \and command from previous versions of AASTeX is now
%% depreciated in this version as it is no longer necessary. AASTeX 
%% automatically takes care of all commas and "and"s between authors names.

%% AASTeX 6.2 has the new \collaboration and \nocollaboration commands to
%% provide the collaboration status of a group of authors. These commands 
%% can be used either before or after the list of corresponding authors. The
%% argument for \collaboration is the collaboration identifier. Authors are
%% encouraged to surround collaboration identifiers with ()s. The 
%% \nocollaboration command takes no argument and exists to indicate that
%% the nearby authors are not part of surrounding collaborations.

%% Mark off the abstract in the ``abstract'' environment. 
\begin{abstract}
Type~I Be/X-ray binary outbursts are driven by mass transfer from a Be star decretion disc to a neutron star companion during each orbital period. \cite{Treiber2021} recently observed non-periodic type~I outbursts in RX~J0529.8-6556 that has unknown binary orbital properties.    We show that non-periodic type~I outbursts may be temporarily driven in a low eccentricity binary with a disc that is inclined sufficiently to be  mildly unstable to Kozai-Lidov oscillations. The inclined disc becomes eccentric and material is transferred to the neutron star at up to three locations in each orbit: when the neutron star passes the disc apastron or one of the two nodes of the disc. The timing and magnitude of each vary with the disc argument of periapsis and longitude of the ascending node that precess in opposite directions.  Calculating the orbital period of the RX~J0529.8-6556 system is non-trivial but we suggest it maybe $>300\,\rm day$, longer than previous estimates.  
\end{abstract}

%% Keywords should appear after the \end{abstract} command. 
%% See the online documentation for the full list of available subject
%% keywords and the rules for their use.
\keywords{accretion, accretion disks - binaries: general - stars: emission-line, Be}

%% From the front matter, we move on to the body of the paper.
%% Sections are demarcated by \section and \subsection, respectively.
%% Observe the use of the LaTeX \label
%% command after the \subsection to give a symbolic KEY to the
%% subsection for cross-referencing in a \ref command.
%% You can use LaTeX's \ref and \label commands to keep track of
%% cross-references to sections, equations, tables, and figures.
%% That way, if you change the order of any elements, LaTeX will
%% automatically renumber them.
%%
%% We recommend that authors also use the natbib \citep
%% and \citet commands to identify citations.  The citations are
%% tied to the reference list via symbolic KEYs. The KEY corresponds
%% to the KEY in the \bibitem in the reference list below. 

\section{Introduction}

Be/X-ray binaries typically  consist of a Be star  in a binary orbit with a neutron  star companion  \citep[e.g.][]{Reig2011,haberl2016}. The Be star is rapidly rotating \citep{Slettebak1982,Porter1996} and a Keplerian decretion disc forms from material that is ejected from the equator \citep{Pringle1991,Lee1991,Hanuschik1996}. X-ray outbursts are driven when material is transferred from the Be star disc to the neutron star. Type~I Be/X-ray binary outbursts are most often observed as periodic outbursts. There are at least two mechanisms for driving type~I outbursts. 

First, and most commonly, if the Be star and the neutron star are in an eccentric orbit, the neutron star is able to capture material close to  each periastron passage \citep{Okazaki2001,Negueruela2001,Okazaki2007}.  This mechanism is only dependent on there being a moderate binary eccentricity. If the disc is coplanar to the binary orbit, outbursts occur once per orbital period and close to each periastron passage. If the disc is slightly inclined, the disc undergoes retrograde nodal precession but a regular pattern of outbursts can still occur on a timescale that is slightly $<P_{\rm orb}$ \citep{Martinetal2014b}.\footnote{See figure 4 in \cite{Martinetal2014b} where the small type~I like outbursts at the beginning of the simulation are occurring on a timescale slightly shorter than the orbital period.} The outburst structure varies depending on the difference between the longitude of ascending node of the disc and the argument of periapsis for the binary orbit. The neutron star may be able to capture material once or twice per binary orbit \citep{Okazaki2013}. If the longitude of ascending node of the disc and the binary eccentricity vector are perpendicular to each other, then the neutron star captures material twice per orbit. On the other hand, if they are aligned to each other, then material is only captured once per orbit. As the disc nodally precesses the outbursts transition between these two states but in both cases the outbursts are periodic. 
 
Second, regular outbursts can occur in circular orbit binaries  if the  disc becomes eccentric \citep{Franchini2019b}.  There are a small class of Be/X-ray binaries that have close to circular orbits and yet show type~I outbursts \citep{Pfahl2002,Reig2007b,Cheng2014}.  For discs that are close to coplanar, disc eccentricity may grow through the 3:1 resonance \citep{Lubow1991,Lubow1991b,Lubow1992}. Since the disc apsidally precesses in a prograde direction, the timescale between outbursts is slightly $>P_{\rm orb}$, but  constant in time \citep[see Fig.~4 in][]{Franchini2019b}. This mechanism operates for low inclination discs, $i \lesssim 20^\circ$.  Above this inclination, the disc eccentricity does not grow through this mechanism since the strength of the resonance is greatly reduced outside of the binary orbital plane. 

A hydrodynamical gas disc that is highly misaligned  can undergo global Kozai-Lidov \citep[KL,][]{Kozai1962,Lidov1962}  oscillations where the disc inclination and eccentricity are exchanged \citep{Martinetal2014b,Fu2015,Fu2015b}. The critical inclination required for disc KL oscillations depends sensitively on the disc aspect ratio \citep{Lubow2017, Zanazzi2017}.   During KL oscillations, the disc overflows its Roche lobe and material is transferred to the companion \citep{Franchini2019}. The more highly inclined the disc, the more eccentric the disc becomes and the more mass that is transferred. Highly eccentric discs may be the cause of type~II outbursts that occur less frequently but are more luminous \citep{Martinetal2014,Martin2019kl}.  
%Type~I outbursts in highly inclined discs have not been considered previously.
 
Recently, \cite{Treiber2021} observed the optical light curve for the system RX J0529.8-6556  and found that the outburst period changes over time. The outbursts are non-periodic as the predicted orbital period changes from about $149\,\rm day$ to $200\,\rm day$ over about 10 years.   Optical and X-ray outbursts are normally thought to coincide, however, with the limited X-ray data available, this does not seem to be the case. This can occur when the disc is viewed edge on and is obscuring the star \citep{Rajoelimanana2011}, however, the  positive correlation between the magnitude and colour indicates a relatively low (non edge-on) inclination angle for RX J0529.8-6556 \citep[see Fig. 10 in ][]{Treiber2021}.  The variation of the outburst period and the out of phase X-ray outbursts may be caused by a misaligned disc that is undergoing nodal precession \citep{Treiber2021}.

Since the observed outbursts are non-periodic this suggests that the binary eccentricity must be small. If the binary eccentricity was large, the outbursts would be close to periodic no matter the disc inclination.  In this Letter, we suggest that there are two requirements for non-periodic outbursts:
\begin{enumerate}
    \item The binary eccentricity is low.
    \item The disc inclination is moderate.
\end{enumerate}
The inclination must be high enough for the disc to be mildly unstable to Kozai-Lidov disc oscillations to become eccentric but not too high that the disc becomes significantly eccentric and undergoes type~II like outbursts.  In Section~\ref{hydro} we show a hydrodynamic simulation of a Be star disc in a configuration that is mildly KL unstable and the disc eccentricity grows to moderate values. We show that we can broadly reproduce the observed non-periodic outbursts and calculate an approximate orbital period. We conclude in Section~\ref{conc}. 
 
\begin{figure*}
\begin{center}
\includegraphics[width=0.45\columnwidth]{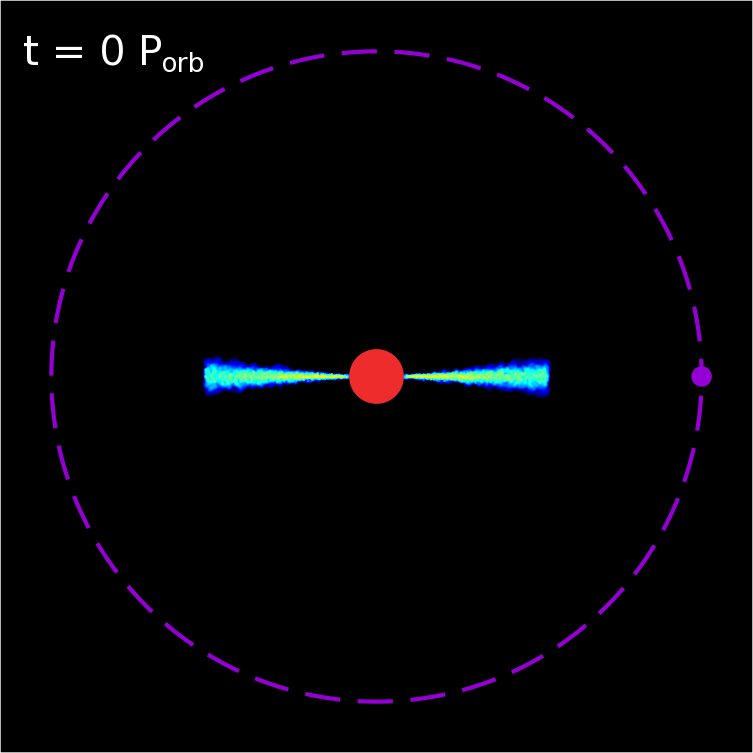}
\includegraphics[width=1.35\columnwidth]{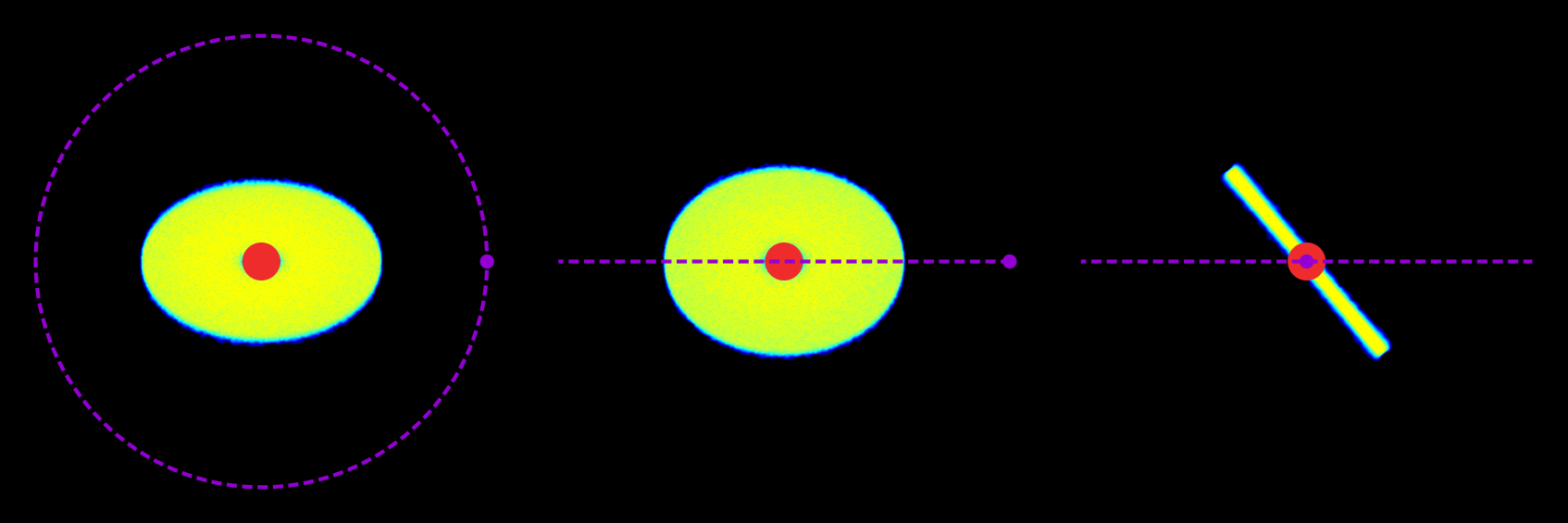}
\includegraphics[width=0.45\columnwidth]{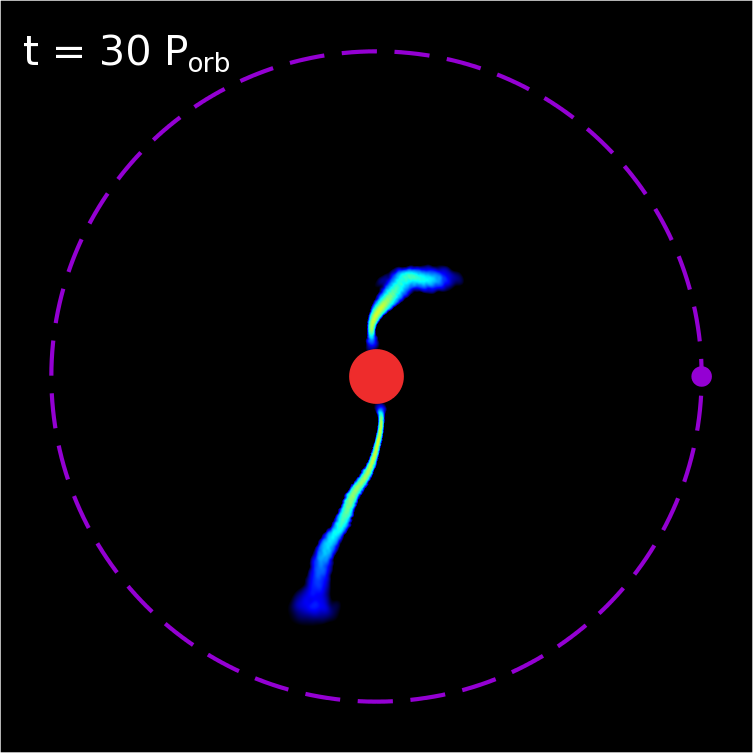}
	\includegraphics[width=1.35\columnwidth]{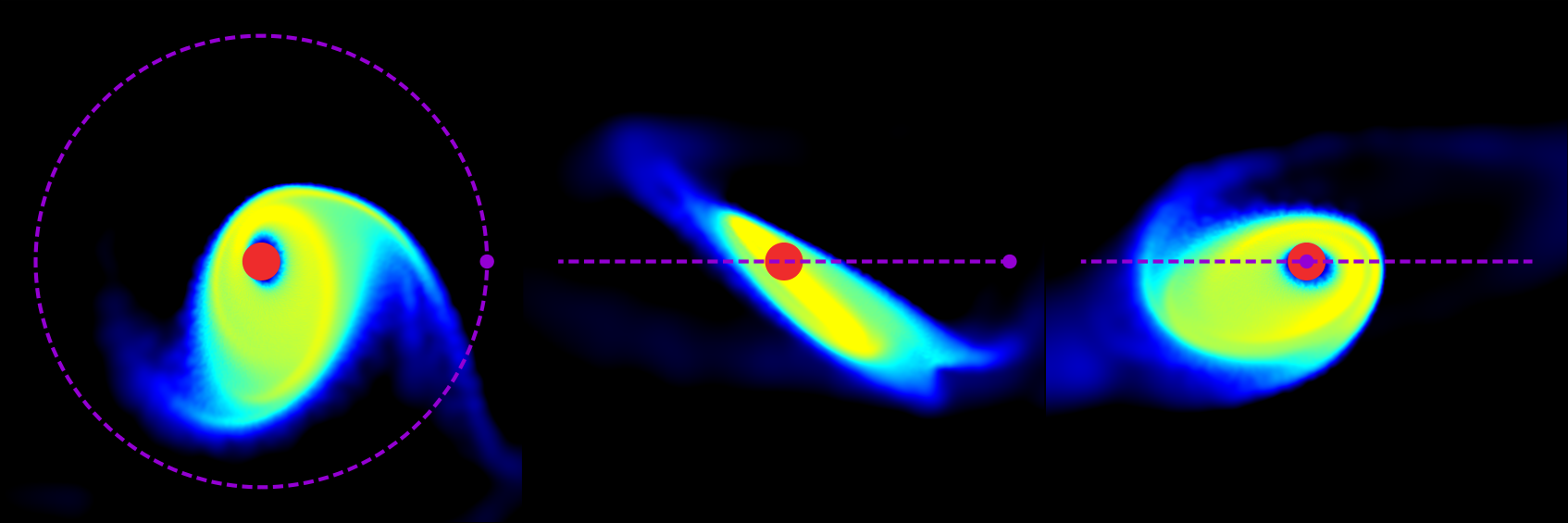}
	\includegraphics[width=0.45\columnwidth]{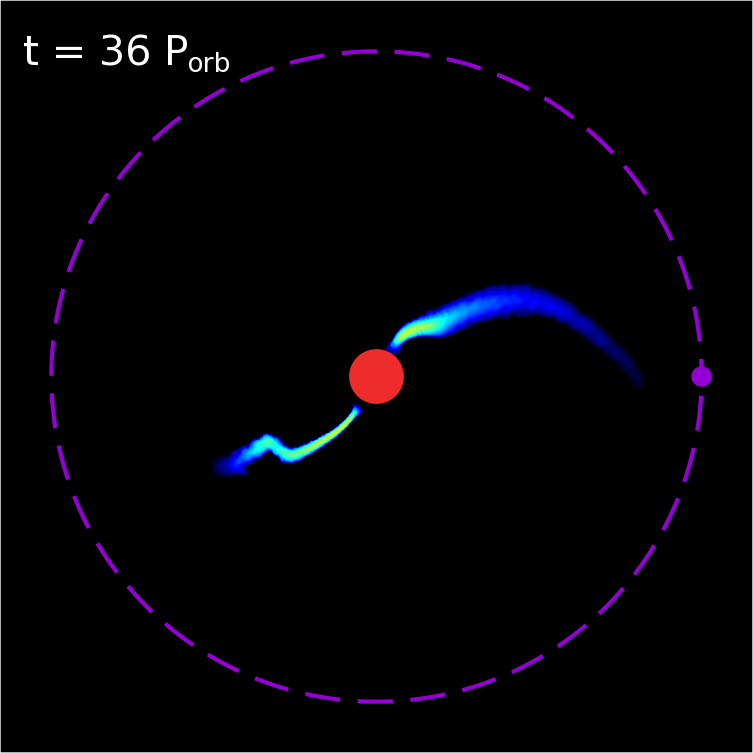}
		\includegraphics[width=1.35\columnwidth]{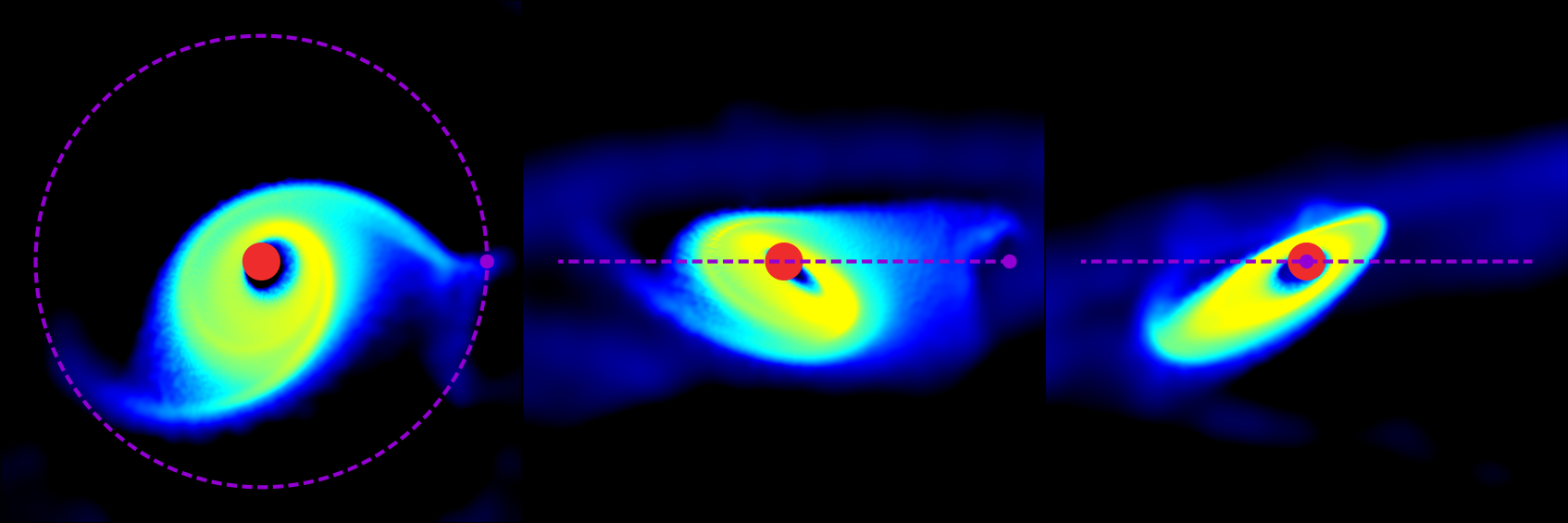}
		\includegraphics[width=0.45\columnwidth]{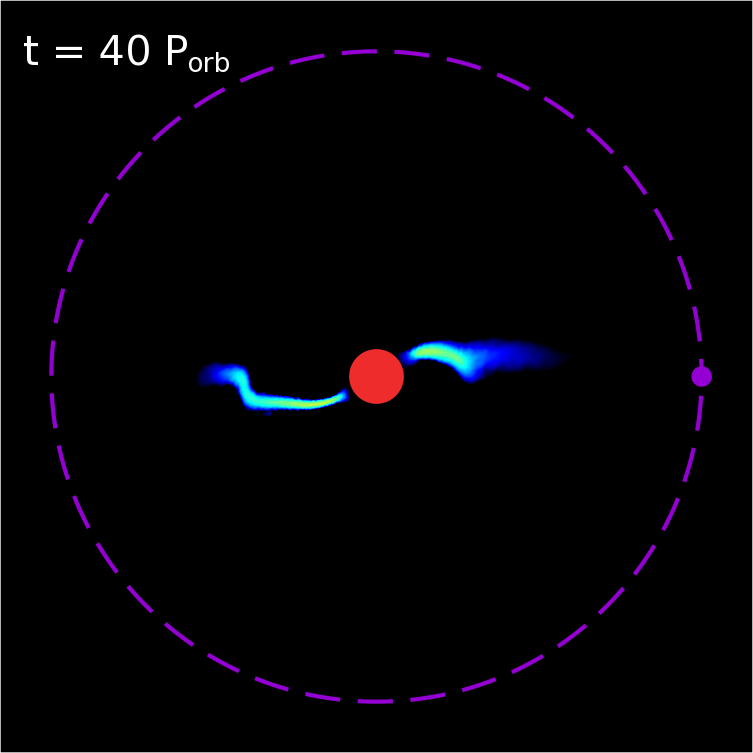}
	\includegraphics[width=1.35\columnwidth]{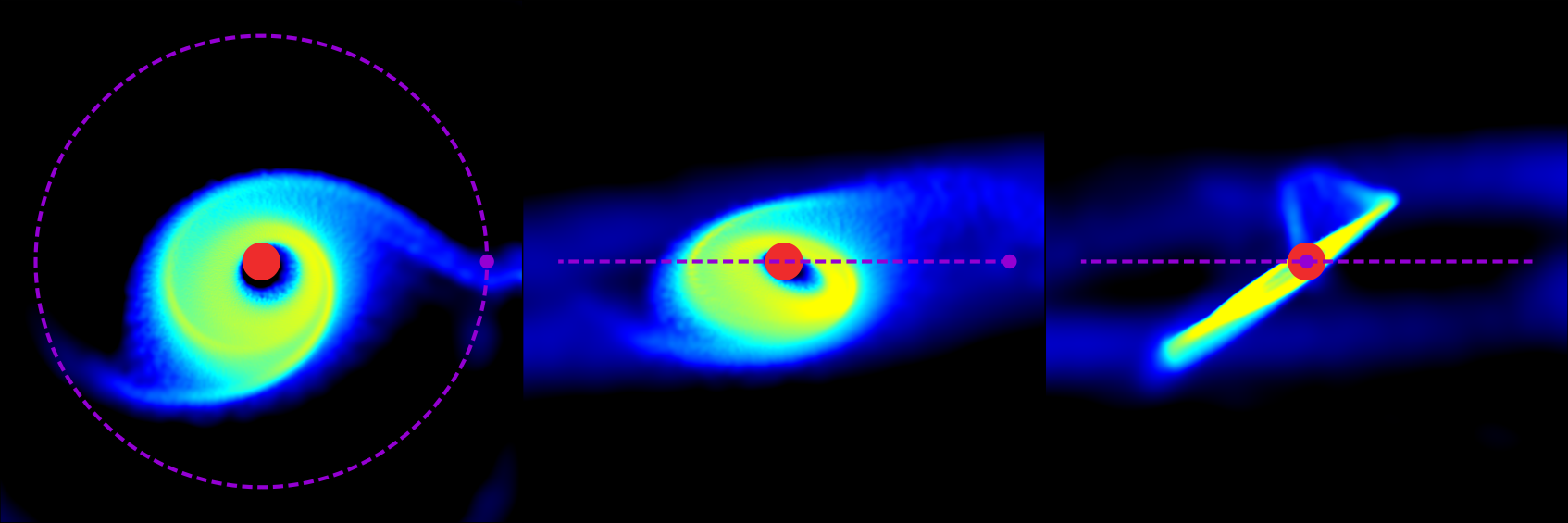}
	\includegraphics[width=0.45\columnwidth]{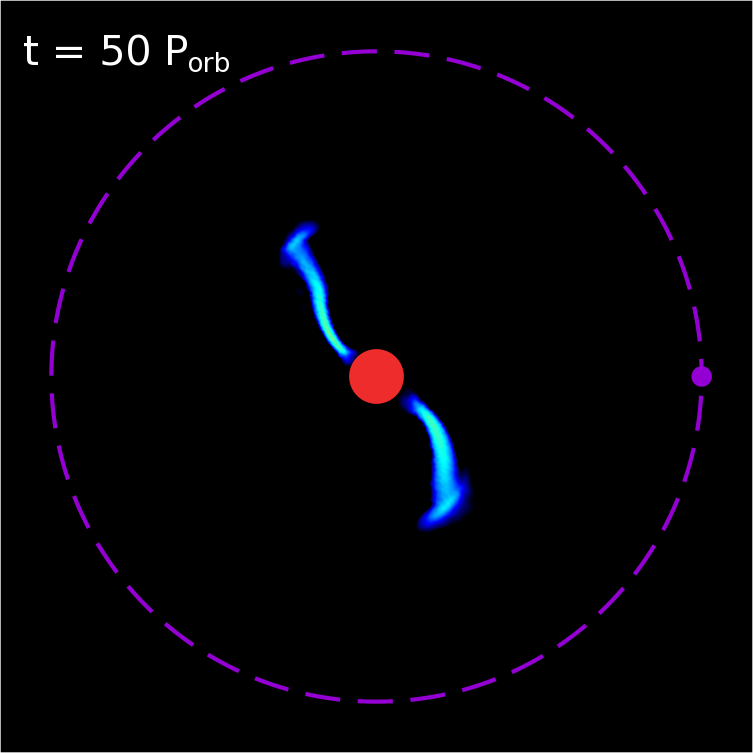} 
	\includegraphics[width=1.35\columnwidth]{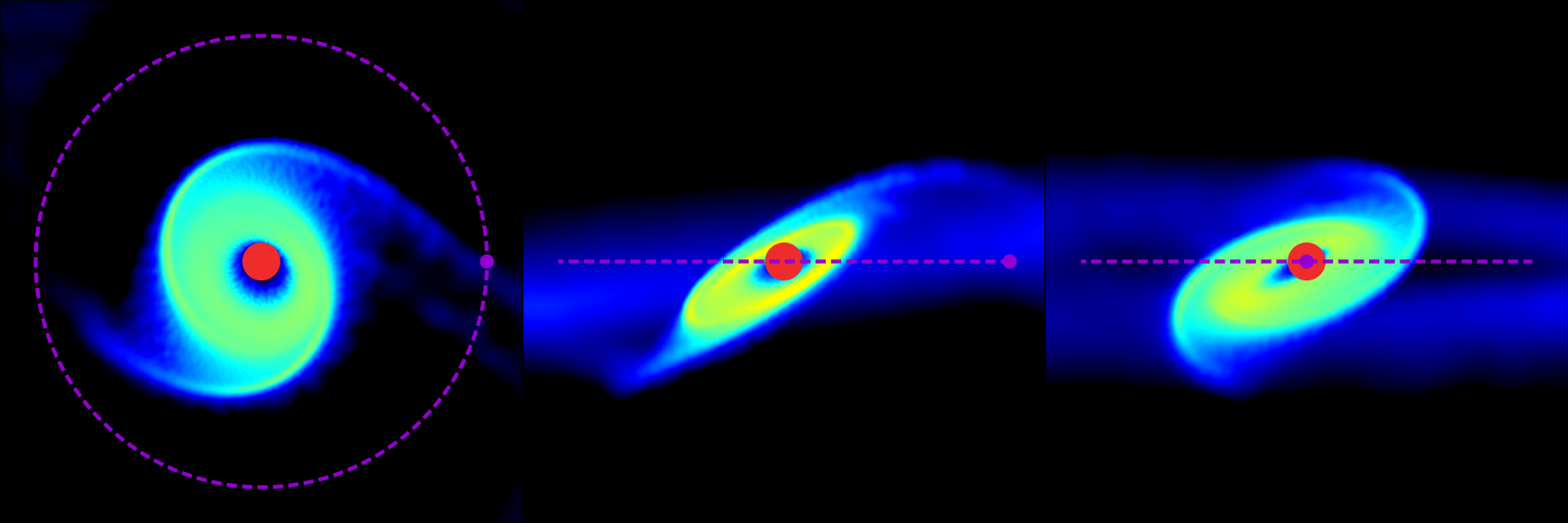} 
	\end{center}
    \caption{Disc evolution in a frame that is corotating with the binary and centered on the Be star. In each row, the left panel shows the cross section of the disc in the binary orbital plane, the $x-y$ plane. The other three plots show the disc with a view in the $x-y$ plane (second column), the $y-z$ plane (third column) and the $x-z$ plane (right column).  The red circles show the Be star scaled to the sink size. The magenta circle shows the neutron star (enlarged from its sink size) and the magenta dashed lines show the neutron star orbit.  The times shown are $t=0\,P_{\rm orb}$ (top), $t=30\,P_{\rm orb}$ (second row), $t=36\,P_{\rm orb}$ (third row),  $t=40\,P_{\rm orb}$ (fourth row)  and $t=50\,P_{\rm orb}$ (bottom row). } 
    \label{splash}
\end{figure*}

\section{Hydrodynamic simulations}
\label{hydro}

\begin{figure}
			\includegraphics[width=\columnwidth]{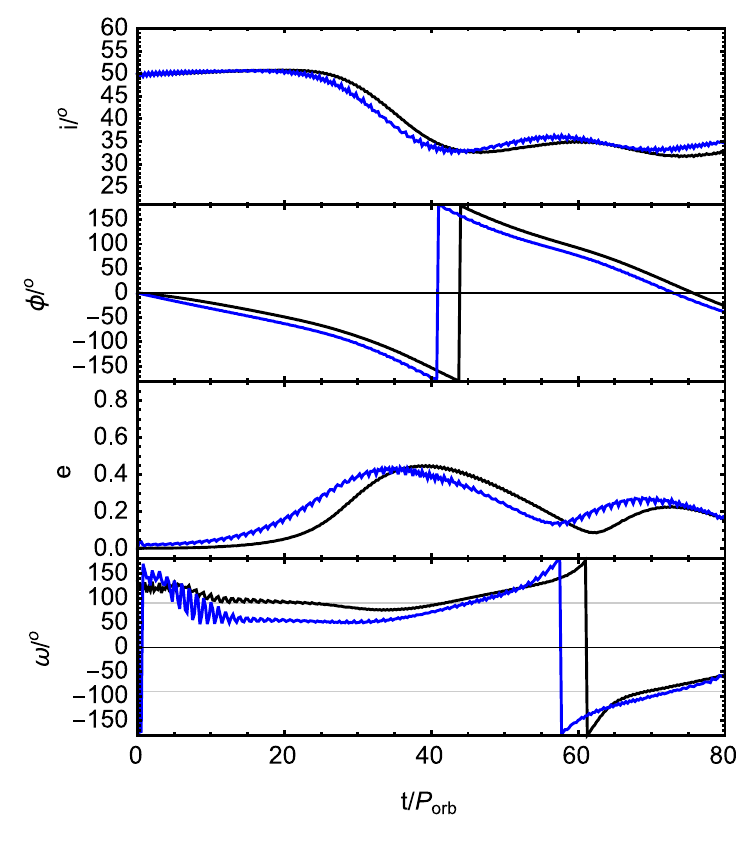} 
    \caption{Evolution of the Be star disc at a semi-major axes of $20\,\rm R_\odot$ (black lines) and $40\,\rm R_\odot$ (blue lines). The  panels show the disc inclination (upper), longitude of ascending node (second),  eccentricity (third) and  argument of periapsis (bottom). The horizontal gray lines in the bottom panel show $\omega=\pm 90^\circ$.} 
    \label{fig:incecc}
\end{figure}
 
There are many parameters of the RX J0529.8-6556 binary system that are unknown. These include the binary inclination, eccentricity and orbital period as well as disc parameters such as the aspect ratio and viscosity.  As a result, we do not try to explore all of the parameter space to find the best fitting model to the observed light curve.  Further we do not extract the exact periodicities from the data but leave that to future work \citep[see for example][]{Feigelson2018,Caceres2019}. Instead, in this Section we simply consider a test case of a hydrodynamical Be star disc simulation to show the principle that non-periodic outbursts are possible in a misaligned viscous disc model.  

\subsection{Simulation set-up}

We use the {\sc Phantom} smoothed particle hydrodynamics (SPH) code \citep{Price2010,Price2018} to model a Be star decretion disc with a neutron star companion on a circular orbit. We use the same disc parameters as \citep{Martinetal2014}. The Be star has a mass of $18\,\rm M_\odot$ and the neutron star has a mass of $1.4\,\rm M_\odot$. The stars are modelled with sink particles. The particles that pass inside the sink radius are accreted and their mass and angular momentum are added to the sink \citep{Bateetal1995}. The sink radius of the Be star is $8\,\rm R_\odot$ and the neutron star is $1\,\rm R_\odot$. The orbit is circular and the orbital period is $24\,\rm day$. We discuss how our results apply to longer orbital period binaries later.

The disc has an initial mass of $M_{\rm d}=10^{-8}\,\rm M_\odot$. This is small enough that it has a negligible effect on the binary orbit and we do not include self-gravity in our calculations. The material is initially distributed with surface density $\Sigma \propto R^{-1}$ between $R_{\rm in}=8\,\rm R_\odot$ up to $R_{\rm out}=50\,\rm R_{\odot}$ with 500,000 SPH particles. The disc is isothermal with an aspect ratio of $H/R=0.01$ at the disc inner edge, ensuring that it is unstable to KL oscillations. The \cite{SS1973} $\alpha$ viscosity parameter is 0.3 \citep[e.g.][]{Jones2008,Carciofi2012,Rimulo2018}. This is typical for a fully ionised disc \citep{Kingetal2007,Martin2019}. The disc viscosity is implemented with the methods described in \cite{Lodato2010} with $\alpha_{\rm AV}=4.7$ and $\beta_{\rm AV}=6$. The disc is resolved with mean smoothing length per scale height of $\left<h\right>/H=0.64$. The disc is inclined by $50^\circ$ to the binary orbital plane initially. The upper row in Fig.~\ref{splash} shows the initial set up. In order to analyse the dynamics of the Be star disc we bin the particles into 300 bins in particle semi-major axis. Within each bin we average the properties of the individual particles.

\subsection{Disc dynamics and outburst locations}

Fig.~\ref{fig:incecc} shows the evolution of the disc inclination, longitude of ascending node, eccentricity and argument of periapsis at semi-major axes of $20\,\rm R_\odot$ and $40\,\rm R_\odot$ as a function of time. The disc behaves in a similar way at different radii. The disc undergoes global damped KL oscillations. The eccentricity of the disc increases but only up to a maximum of about 0.4. This is not sufficient to drive type~II like outbursts but still has a significant effect on the smaller type~I outbursts.

There are three possible locations where the neutron star can accrete material during each orbit and therefore three possible outburst locations. These are where the the neutron star passes  $\Omega$ (longitude of the ascending node of the disc), $\Omega+\pi$ (longitude of the descending node) and $\Omega+\omega$ (disc apastron). We calculate the precession timescales with $P_{\rm node}=2\pi/\dot{\Omega}$ (nodal precession timescale) and $P_{\rm apastron}=2\pi/(\dot{\omega}+\dot{\Omega})$ (disc apastron precession timescale). Then the timescale between consecutive passages is calculated with
\begin{equation}
    P_{\rm burst}=P_{\rm orb}\left( 1+ \frac{P_{\rm orb}}{P_{\rm prec}} \right)
    \label{burst}
\end{equation}
\citep[e.g.][]{Whitehurst1988}, where $P_{\rm prec}$ is equal to $P_{\rm node}$ or $P_{\rm apastron}$ depending on the location of the outburst.

\begin{figure}
\centering
			\includegraphics[width=\columnwidth]{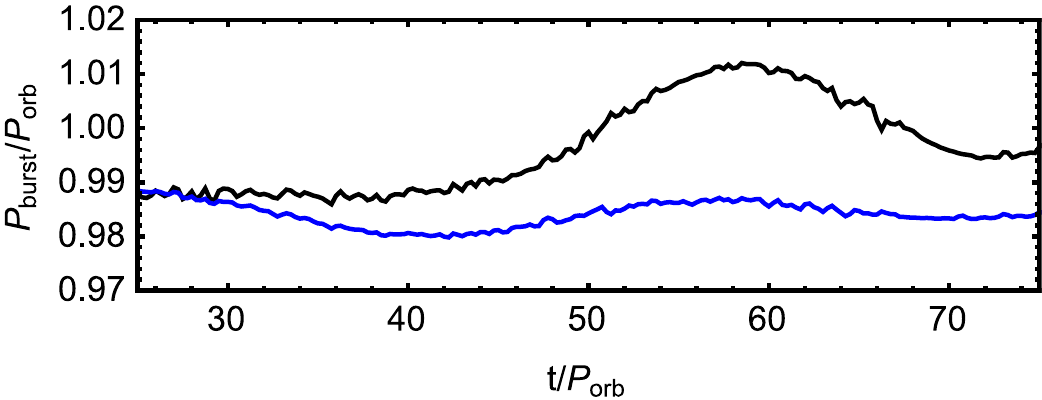} 
    \caption{The timescale between the neutron star passing the disc longitude of ascending node (blue) and the disc apastron (black) calculated with equation~(\ref{burst}).  } 
    \label{pburst}
\end{figure}

Fig.~\ref{pburst} shows the timescale of the neutron star passing the disc longitude of ascending node (blue) and the disc apastron (black).  There are two competing precessions. The nodal precession is retrograde and so the timescale between the node crossings is always slightly less then $P_{\rm orb}$  while the apsidal precession is always prograde. The time between disc apastron crossing depends on the balance between the nodal precession and the apsidal precession. Initially the disc apastron crossing timescale is shorter than the orbital period, but after about $t=50\,P_{\rm orb}$ it becomes longer.  This clearly shows that the burst timescales are different for these two locations. The number of outbursts each orbit depends upon the argument of periapsis for the disc. If the argument of periapsis is aligned to a node of the disc ($\omega \approx 0^\circ$ or $180^\circ$), there can be only one outburst per orbital period. If $\omega\approx \pm 90^\circ$, there can be three outbursts per orbit. In between there may be two outbursts. Thus, a misaligned and eccentric disc can have between one and three bursts per orbit with varying magnitudes.

\subsection{Accretion rate on to the neutron star}

\begin{figure*}
\centering
		\includegraphics[width=2\columnwidth]{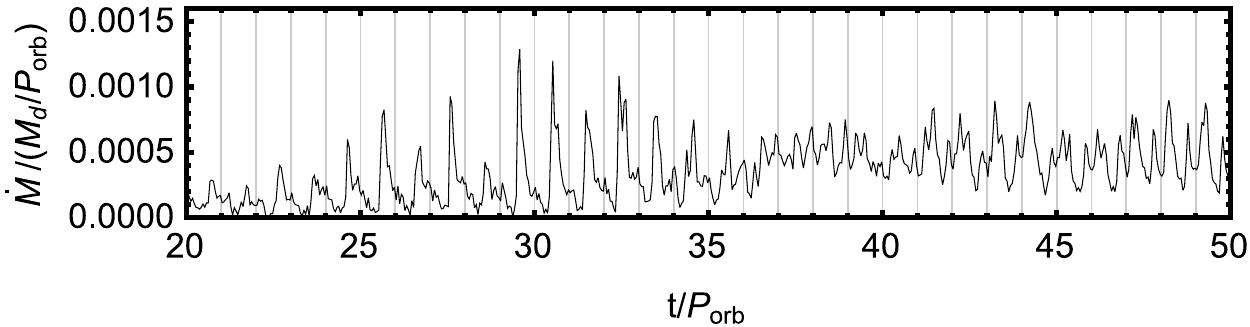} 
			\includegraphics[width=2\columnwidth]{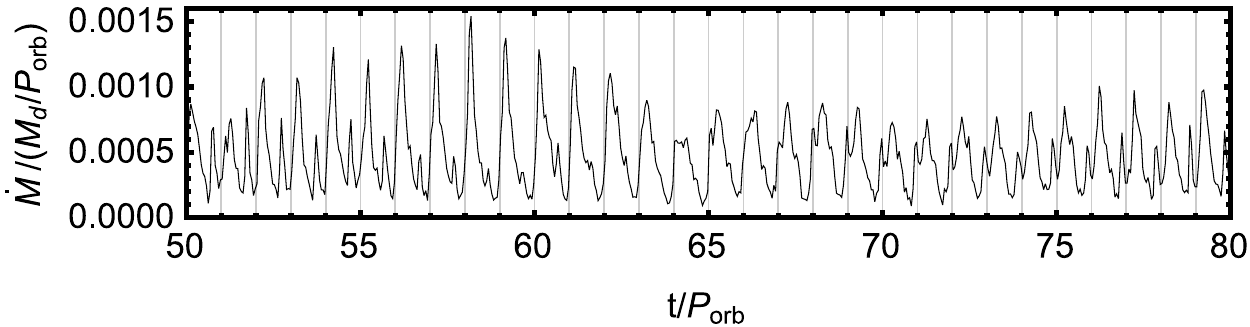} 
    \caption{Accretion rate on to the neutron star as a function of time. } 
    \label{treiber}
\end{figure*}
 
Fig.~\ref{treiber}  shows the accretion rate on to the neutron star as a function of time. We do not show the early evolution before $t=20\,P_{\rm orb}$ since initially there is little flow as the disc expands outwards. A misaligned disc has a larger tidal truncation radius than a coplanar disc \citep{Lubowetal2015,Miranda2015}. The disc eccentricity grows due to the KL effect  and by a time of about $20\,P_{\rm orb}$ outbursts occur once per orbital period.  These outbursts are a result of the disc eccentricity and they occur when the neutron star passes close to the disc apastron.  In the second row of Fig.~\ref{splash} we show the disc at time $t=30\,P_{\rm orb}$.  The cross section of the disc in the binary orbital plane (left) shows that  the disc is much more extended on the lower side than the upper side and this explains why there is only one outburst per orbital period. 
The argument of periapsis of the disc is about $50^\circ$ (at a semi-major axis of $40\,\rm R_\odot$, see the blue line in the bottom row of Fig. \ref{fig:incecc}), this is somewhat in between the criteria for clear one or three outbursts per orbital period. However, as we can see from Fig. \ref{treiber}, the outburst is dominated by the disc apastron passage and there is therefore only one peak. The subpeaks start to become evident at roughly $34\,P_{\rm orb}$ as the argument of periapsis moves closer to $90^\circ$.
Note that at this time the two competing precessions give a similar $P_{\rm burst}/P_{\rm orb}$ (see Fig. \ref{pburst}) and this is $<1$.

The third row in Fig.~\ref{splash} shows the disc at a time of $t=36\,P_{\rm orb}$. The disc eccentricity is close to its maximum value of around 0.4.  However, the cross section shows that the disc is smaller in the binary orbital plane than at $t=30\,P_{\rm orb}$ because the disc has apsidally precessed. The accretion is dominated by that from the node crossings and the outbursts occur on a timescale slightly less than $P_{\rm orb}/2$ between $t=36-39\,\rm P_{\rm orb}$.  During this period of high eccentricity,  the average accretion rate on to the neutron star is much higher  but the amplitude of these outbursts is relatively low.

At a time of about $40\,P_{\rm orb}$ the accretion  again has one main outburst per orbital period but  sub peaks can be seen.  The fourth row in Fig.~\ref{splash} shows the disc at this time. The argument of periapsis is $81^\circ$ and the size of the disc in the cross section of the binary orbital plane is small. The amplitude of the outbursts grows in time for a few orbital periods. At a time of about $t=45\,P_{\rm orb}$, the argument of periapsis passes through $90^\circ$. At this time we clearly see the three peaks per orbital period. The middle and largest amplitude peak is due to the disc eccentricity while the subpeaks on each side are where the neutron star passes the nodes of the disc.  The delay between the disc eccentricity peak and the following  node peak is particularly evident at $41-45\,P_{\rm orb}$ in Fig.~\ref{treiber}.

The bottom row in Fig.~\ref{splash} shows the disc at a time of $t=50\,P_{\rm orb}$ where the disc eccentricity has significantly decreased. The eccentricity at this time is similar to that at $t=30\,P_{\rm orb}$, however, the accretion rate shows two outbursts per orbital period compared to one at $t=30\,P_{\rm orb}$. The cross section in the binary orbital plane shows that the disc is fairly symmetric in the this plane. The outbursts occur when the neutron star passes close to the two nodes of the  disc on each orbital period and the outbursts are of similar magnitudes. 
We note that in the right hand panels of Fig.~\ref{splash} we can see that there is a circumbinary disc forming around the binary \citep[see also][]{Franchini2019b}.

At a time of around $t=72\,P_{\rm orb}$ the disc reaches the peak eccentricity of the second KL cycle (see  Fig.~\ref{fig:incecc}). The argument of periapsis passes through $-90^\circ$ and we again see evidence of three outbursts per orbital period with the largest magnitude outburst coming from the disc apastron crossing. The material captured at the disc apastron dominates and we do not get the non-periodic behaviour again. 
We have not included accretion into the Be star disc from the star and so the inclination of the disc decays over time.  However, if there was a source of high inclination material being added to the disc the KL disc oscillations may be more long lived. This has been seen in the case of a circumbinary disc that feeds the formation of circumstellar discs at high inclination \citep{Smallwood2021}. Thus, the non-periodic outburst phase may be able to repeat in time and the outbursts would periodically show these frequency changes \citep[see also][]{Suffak2021}. 

\subsection{Application to RX~J0529.8-6556}

The orbital period of the binary that we have considered in the simulation is relatively short. However, we expect the physics to be the same for a system with longer orbital period provided that the disc is sufficiently large to be tidally truncated \citep[e.g.][]{martin2011be}.  We have also run a simulation with a small binary eccentricity of $0.1$ and we find very similar behaviour.  Bearing these in mind, we can compare our model to the observed lightcurve for RX~J0529.8-6556. The accretion rate in  Fig.~\ref{treiber} between a time of about $37-48\,P_{\rm orb}$ has a similar shape to the lightcurve observed in \cite{Treiber2021}. Initially the outbursts occur on a short timescale that is about $P_{\rm orb}/2$ and these outbursts have a relatively small amplitude. The time between outbursts increases over time as does the amplitude of the outbursts. While we have not tried to tune our model parameters to get the closest possible fit, we can make a prediction about the orbital period of the system. The short period outbursts at the start ($t\approx 37\,P_{\rm orb}$) are on a timescale of $0.98\,P_{\rm orb}/2$ (see the blue line in Fig.~\ref{pburst}). In the observed light curve, the outbursts are every $149\,\rm day$ and so we suggest that the orbital period of this system could be about $305\,\rm day$.

We also ran a simulation with the same initial disc inclination, i.e. $i=50^\circ$, but with a binary eccentricity of $0.34$. While the disc has low eccentricity, the dominant effect is the binary eccentricity and outbursts occur close to the binary periastron passage on the orbital period.  When the disc becomes eccentric because of the KL effect, the outbursts are dominated by the disc eccentricity and these occur once per orbital period when the neutron star passes the disc apastron. The transition between these two regimes is quick and we do not observe significant changes to the outburst period aside from at these transitions.

\section{Conclusions}
\label{conc}

Periodic type I outbursts may be driven by either a moderate binary orbital eccentricity or disc eccentricity in a circular orbit binary. We have shown that non-periodic outbursts may be temporarily driven in a binary that is close to circular with an inclination such that the disc is mildly unstable to KL disc oscillations. The inclined and eccentric disc transfers material at up to three locations during each orbital period: the two nodes of the disc and the disc apastron.  The relative magnitude of each depends upon the  argument of periapsis of the disc. The timing of each outburst depends upon the rate of the nodal and apsidal precessions that go in opposite directions. The rapid nodal precession during a KL cycle can lead to rapid changes to the outburst period.

\begin{acknowledgements}

We thank an anonymous referee for useful comments. Computer support was provided by UNLV’s National Supercomputing Center.  RGM acknowledges support from NASA through grant 80NSSC21K0395.
AF acknowledges financial support provided under the European Union's H2020 ERC Consolidator Grant ``Binary Massive Black Hole Astrophysics" (B Massive, Grant Agreement: 818691).
We acknowledge the use of SPLASH \citep{Price2007} for the rendering of Fig.~\ref{splash}.

\end{acknowledgements}

%% The reference list follows the main body and any appendices.
%% Use LaTeX's thebibliography environment to mark up your reference list.
%% Note \begin{thebibliography} is followed by an empty set of
%% curly braces.  If you forget this, LaTeX will generate the error
%% "Perhaps a missing \item?".
%%
%% thebibliography produces citations in the text using \bibitem-\cite
%% cross-referencing. Each reference is preceded by a
%% \bibitem command that defines in curly braces the KEY that corresponds
%% to the KEY in the \cite commands (see the first section above).
%% Make sure that you provide a unique KEY for every \bibitem or else the
%% paper will not LaTeX. The square brackets should contain
%% the citation text that LaTeX will insert in
%% place of the \cite commands.

%% We have used macros to produce journal name abbreviations.
%% \aastex provides a number of these for the more frequently-cited journals.
%% See the Author Guide for a list of them.

%% Note that the style of the \bibitem labels (in []) is slightly
%% different from previous examples.  The natbib system solves a host
%% of citation expression problems, but it is necessary to clearly
%% delimit the year from the author name used in the citation.
%% See the natbib documentation for more details and options.

\bibliographystyle{aasjournal}
%%%%%\bibliography{martin} % if your bibtex file is called example.bib

\begin{thebibliography}{}
\expandafter\ifx\csname natexlab\endcsname\relax\def\natexlab#1{#1}\fi
\providecommand{\url}[1]{\href{#1}{#1}}

\bibitem[{{Bate} {et~al.}(1995){Bate}, {Bonnell}, \& {Price}}]{Bateetal1995}
{Bate}, M.~R., {Bonnell}, I.~A., \& {Price}, N.~M. 1995, \mnras, 277, 362

\bibitem[{{Caceres} {et~al.}(2019){Caceres}, {Feigelson}, {Jogesh Babu},
  {Bahamonde}, {Christen}, {Bertin}, {Meza}, \& {Cur{\'e}}}]{Caceres2019}
{Caceres}, G.~A., {Feigelson}, E.~D., {Jogesh Babu}, G., {et~al.} 2019, \aj,
  158, 58

\bibitem[{{Carciofi} {et~al.}(2012){Carciofi}, {Bjorkman}, {Otero}, {Okazaki},
  {{\v S}tefl}, {Rivinius}, {Baade}, \& {Haubois}}]{Carciofi2012}
{Carciofi}, A.~C., {Bjorkman}, J.~E., {Otero}, S.~A., {et~al.} 2012, \apjl,
  744, L15

\bibitem[{{Cheng} {et~al.}(2014){Cheng}, {Shao}, \& {Li}}]{Cheng2014}
{Cheng}, Z.~Q., {Shao}, Y., \& {Li}, X.~D. 2014, \apj, 786, 128

\bibitem[{{Feigelson} {et~al.}(2018){Feigelson}, {Babu}, \&
  {Caceres}}]{Feigelson2018}
{Feigelson}, E.~D., {Babu}, G.~J., \& {Caceres}, G.~A. 2018, Frontiers in
  Physics, 6, 80

\bibitem[{{Franchini} \& {Martin}(2019)}]{Franchini2019b}
{Franchini}, A., \& {Martin}, R.~G. 2019, \apjl, 881, L32

\bibitem[{{Franchini} {et~al.}(2019){Franchini}, {Martin}, \&
  {Lubow}}]{Franchini2019}
{Franchini}, A., {Martin}, R.~G., \& {Lubow}, S.~H. 2019, \mnras, 485, 315

\bibitem[{{Fu} {et~al.}(2015{\natexlab{a}}){Fu}, {Lubow}, \& {Martin}}]{Fu2015}
{Fu}, W., {Lubow}, S.~H., \& {Martin}, R.~G. 2015{\natexlab{a}}, \apj, 807, 75

\bibitem[{{Fu} {et~al.}(2015{\natexlab{b}}){Fu}, {Lubow}, \&
  {Martin}}]{Fu2015b}
---. 2015{\natexlab{b}}, ApJ, 813, 105

\bibitem[{{Haberl} \& {Sturm}(2016)}]{haberl2016}
{Haberl}, F., \& {Sturm}, R. 2016, \aap, 586, A81

\bibitem[{{Hanuschik}(1996)}]{Hanuschik1996}
{Hanuschik}, R.~W. 1996, A\&A, 308, 170

\bibitem[{{Jones} {et~al.}(2008){Jones}, {Sigut}, \& {Porter}}]{Jones2008}
{Jones}, C.~E., {Sigut}, T.~A.~A., \& {Porter}, J.~M. 2008, \mnras, 386, 1922

\bibitem[{{King} {et~al.}(2007){King}, {Pringle}, \& {Livio}}]{Kingetal2007}
{King}, A.~R., {Pringle}, J.~E., \& {Livio}, M. 2007, MNRAS, 376, 1740

\bibitem[{{Kozai}(1962)}]{Kozai1962}
{Kozai}, Y. 1962, AJ, 67, 591

\bibitem[{{Lee} {et~al.}(1991){Lee}, {Osaki}, \& {Saio}}]{Lee1991}
{Lee}, U., {Osaki}, Y., \& {Saio}, H. 1991, MNRAS, 250, 432

\bibitem[{{Lidov}(1962)}]{Lidov1962}
{Lidov}, M.~L. 1962, Planet. Space Sci., 9, 719

\bibitem[{{Lodato} \& {Price}(2010)}]{Lodato2010}
{Lodato}, G., \& {Price}, D.~J. 2010, \mnras, 405, 1212

\bibitem[{{Lubow}(1991{\natexlab{a}})}]{Lubow1991}
{Lubow}, S.~H. 1991{\natexlab{a}}, ApJ, 381, 259

\bibitem[{{Lubow}(1991{\natexlab{b}})}]{Lubow1991b}
---. 1991{\natexlab{b}}, ApJ, 381, 268

\bibitem[{{Lubow}(1992)}]{Lubow1992}
---. 1992, ApJ, 401, 317

\bibitem[{{Lubow} {et~al.}(2015){Lubow}, {Martin}, \& {Nixon}}]{Lubowetal2015}
{Lubow}, S.~H., {Martin}, R.~G., \& {Nixon}, C. 2015, ApJ, 800, 96

\bibitem[{{Lubow} \& {Ogilvie}(2017)}]{Lubow2017}
{Lubow}, S.~H., \& {Ogilvie}, G.~I. 2017, \mnras, 469, 4292

\bibitem[{{Martin} \& {Franchini}(2019)}]{Martin2019kl}
{Martin}, R.~G., \& {Franchini}, A. 2019, \mnras, 489, 1797

\bibitem[{{Martin} {et~al.}(2014{\natexlab{a}}){Martin}, {Nixon}, {Armitage},
  {Lubow}, \& {Price}}]{Martinetal2014}
{Martin}, R.~G., {Nixon}, C., {Armitage}, P.~J., {Lubow}, S.~H., \& {Price},
  D.~J. 2014{\natexlab{a}}, ApJL, 790, L34

\bibitem[{{Martin} {et~al.}(2014{\natexlab{b}}){Martin}, {Nixon}, {Lubow},
  {Armitage}, {Price}, {Do{\u g}an}, \& {King}}]{Martinetal2014b}
{Martin}, R.~G., {Nixon}, C., {Lubow}, S.~H., {et~al.} 2014{\natexlab{b}},
  ApJL, 792, L33

\bibitem[{{Martin} {et~al.}(2019){Martin}, {Nixon}, {Pringle}, \&
  {Livio}}]{Martin2019}
{Martin}, R.~G., {Nixon}, C.~J., {Pringle}, J.~E., \& {Livio}, M. 2019, New
  Astronomy, 70, 7

\bibitem[{{Martin} {et~al.}(2011){Martin}, {Pringle}, {Tout}, \&
  {Lubow}}]{martin2011be}
{Martin}, R.~G., {Pringle}, J.~E., {Tout}, C.~A., \& {Lubow}, S.~H. 2011,
  \mnras, 416, 2827

\bibitem[{{Miranda} \& {Lai}(2015)}]{Miranda2015}
{Miranda}, R., \& {Lai}, D. 2015, \mnras, 452, 2396

\bibitem[{{Negueruela} {et~al.}(2001){Negueruela}, {Okazaki}, {Fabregat},
  {Coe}, {Munari}, \& {Tomov}}]{Negueruela2001}
{Negueruela}, I., {Okazaki}, A.~T., {Fabregat}, J., {et~al.} 2001, A\&A, 369,
  117

\bibitem[{{Okazaki}(2007)}]{Okazaki2007}
{Okazaki}, A.~T. 2007, in Astronomical Society of the Pacific Conference
  Series, Vol. 367, Massive Stars in Interactive Binaries, ed. N.~{St.-Louis}
  \& A.~F.~J. {Moffat}, 485

\bibitem[{{Okazaki} {et~al.}(2013){Okazaki}, {Hayasaki}, \&
  {Moritani}}]{Okazaki2013}
{Okazaki}, A.~T., {Hayasaki}, K., \& {Moritani}, Y. 2013, \pasj, 65, 41

\bibitem[{{Okazaki} \& {Negueruela}(2001)}]{Okazaki2001}
{Okazaki}, A.~T., \& {Negueruela}, I. 2001, A\&A, 377, 161

\bibitem[{{Pfahl} {et~al.}(2002){Pfahl}, {Rappaport}, {Podsiadlowski}, \&
  {Spruit}}]{Pfahl2002}
{Pfahl}, E., {Rappaport}, S., {Podsiadlowski}, P., \& {Spruit}, H. 2002, \apj,
  574, 364

\bibitem[{{Porter}(1996)}]{Porter1996}
{Porter}, J.~M. 1996, MNRAS, 280, L31

\bibitem[{{Price}(2007)}]{Price2007}
{Price}, D.~J. 2007, Pasa, 24, 159

\bibitem[{{Price} \& {Federrath}(2010)}]{Price2010}
{Price}, D.~J., \& {Federrath}, C. 2010, \mnras, 406, 1659

\bibitem[{{Price} {et~al.}(2018){Price}, {Wurster}, {Tricco}, {Nixon},
  {Toupin}, {Pettitt}, {Chan}, {Mentiplay}, {Laibe}, {Glover}, {Dobbs},
  {Nealon}, {Liptai}, {Worpel}, {Bonnerot}, {Dipierro}, {Ballabio}, {Ragusa},
  {Federrath}, {Iaconi}, {Reichardt}, {Forgan}, {Hutchison}, {Constantino},
  {Ayliffe}, {Hirsh}, \& {Lodato}}]{Price2018}
{Price}, D.~J., {Wurster}, J., {Tricco}, T.~S., {et~al.} 2018, \pasa, 35, e031

\bibitem[{{Pringle}(1991)}]{Pringle1991}
{Pringle}, J.~E. 1991, MNRAS, 248, 754

\bibitem[{{Rajoelimanana} {et~al.}(2011){Rajoelimanana}, {Charles}, \&
  {Udalski}}]{Rajoelimanana2011}
{Rajoelimanana}, A.~F., {Charles}, P.~A., \& {Udalski}, A. 2011, \mnras, 413,
  1600

\bibitem[{{Reig}(2007)}]{Reig2007b}
{Reig}, P. 2007, \mnras, 377, 867

\bibitem[{{Reig}(2011)}]{Reig2011}
---. 2011, \apss, 332, 1

\bibitem[{{R{\'{\i}}mulo} {et~al.}(2018){R{\'{\i}}mulo}, {Carciofi}, {Vieira},
  {Rivinius}, {Faes}, {Figueiredo}, {Bjorkman}, {Georgy}, {Ghoreyshi}, \&
  {Soszy{\'n}ski}}]{Rimulo2018}
{R{\'{\i}}mulo}, L.~R., {Carciofi}, A.~C., {Vieira}, R.~G., {et~al.} 2018,
  \mnras, 476, 3555

\bibitem[{{Shakura} \& {Sunyaev}(1973)}]{SS1973}
{Shakura}, N.~I., \& {Sunyaev}, R.~A. 1973, A\&A, 24, 337

\bibitem[{{Slettebak}(1982)}]{Slettebak1982}
{Slettebak}, A. 1982, ApJs, 50, 55

\bibitem[{{Smallwood} {et~al.}(2021){Smallwood}, {Martin}, \&
  {Lubow}}]{Smallwood2021}
{Smallwood}, J.~L., {Martin}, R.~G., \& {Lubow}, S.~H. 2021, \apjl, 907, L14

\bibitem[{{Suffak} {et~al.}(2021){Suffak}, {Jones}, \& {Carciofi}}]{Suffak2021}
{Suffak}, M., {Jones}, C.~E., \& {Carciofi}, A.~C. 2021, arXiv e-prints,
  arXiv:2110.08344

\bibitem[{{Treiber} {et~al.}(2021){Treiber}, {Vasilopoulos}, {Bailyn},
  {Haberl}, {Gendreau}, {Ray}, {Maitra}, {Maggi}, {Jaisawal}, {Udalski},
  {Wilms}, {Monageng}, {Buckley}, {K{\"o}nig}, \& {Carpano}}]{Treiber2021}
{Treiber}, H., {Vasilopoulos}, G., {Bailyn}, C.~D., {et~al.} 2021, \mnras, 503,
  6187

\bibitem[{{Whitehurst}(1988)}]{Whitehurst1988}
{Whitehurst}, R. 1988, \mnras, 232, 35

\bibitem[{{Zanazzi} \& {Lai}(2017)}]{Zanazzi2017}
{Zanazzi}, J.~J., \& {Lai}, D. 2017, \mnras, 467, 1957

\end{thebibliography}

%% This command is needed to show the entire author+affilation list when
%% the collaboration and author truncation commands are used.  It has to
%% go at the end of the manuscript.
%\allauthors

%% Include this line if you are using the \added, \replaced, \deleted
%% commands to see a summary list of all changes at the end of the article.
%\listofchanges

% End of file `sample62.tex'.

\end{document}